\documentclass{article}
\usepackage{hyperref}
\usepackage{graphicx}
\usepackage{amsmath, amssymb, amsthm}
\usepackage{amsfonts}
\usepackage{algorithm}
\usepackage{algorithmicx}
\usepackage{algpseudocode} 
\usepackage{multirow}
\usepackage{tabularx}
\usepackage{url}
\usepackage{booktabs}

\newtheorem{theorem}{Theorem}
\theoremstyle{definition}

\title{Localized Conformal Multi-Quantile Regression}
\author{Yuan Lu\\ 
\small{Department of Biostatistics, Georgetown University}\\
\small{\href{mailto:yl1939@georgetown.edu}{yl1939@georgetown.edu}}}
\date{October, 2025}

\begin{document}

\maketitle

\begin{abstract}
Standard conformal prediction methods guarantee marginal coverage but often produce inefficient intervals that fail to adapt to local heteroscedasticity, while recent localized approaches often struggle to maintain validity across distinct subpopulations with varying noise profiles. To address these challenges, we introduce Localized Conformal Multi-Quantile Regression (LCMQR), a novel framework that synergizes multi-quantile information with kernel-based localization to construct efficient and adaptive prediction intervals. Theoretically, we resolve an inconsistency in Conformalized Composite Quantile Regression (CCQR) by proving that our consistent Average-then-Max scoring mechanism systematically yields tighter intervals than the Max-then-Average approach used in prior work. For heterogeneous environments, we extend this framework to Group-Calibrated LCMQR (GC-LCMQR) via a stratified calibration step that guarantees finite-sample validity within distinct subgroups. Experiments on benchmark datasets and an Individual Treatment Effect (ITE) task demonstrate that LCMQR achieves superior efficiency on standard benchmarks, while GC-LCMQR uniquely achieves group-level coverage for target subgroups in mixture populations where baselines fail.
\end{abstract}

\section{Introduction}

In high-stakes fields such as healthcare, economics, and public policy, providing reliable uncertainty quantification is as critical as accurate point prediction. Quantile Regression (QR) has become a standard tool for this purpose, estimating the conditional quantiles of the response variable to describe predictive uncertainty. However, while standard QR offers a granular view of the conditional distribution, it generally lacks rigorous statistical guarantees for coverage in finite samples. To address this, Conformal Prediction (CP) has emerged as a powerful, model-agnostic framework. Relying solely on the assumption of data exchangeability, CP constructs prediction intervals with valid marginal coverage guarantees, ensuring that the true outcome falls within the predicted interval with a user-specified probability of at least $1-\alpha$.

Building on these foundations, Conformalized Quantile Regression (CQR) combines the strengths of QR with the rigorous guarantees of CP. While CQR has established itself as a robust baseline, it suffers from two significant limitations. First, CQR typically relies on estimating extreme quantiles (e.g., $\alpha/2$ and $1-\alpha/2$) to form intervals. In regimes with limited data or outliers, these estimates exhibit high variance, resulting in unnecessarily wide and unstable intervals. Recent attempts like Conformalized Composite Quantile Regression (CCQR) seek to mitigate this by averaging across multiple quantile levels; however, we identify that CCQR employs a theoretically inconsistent "Max-then-Average" scoring mechanism that creates a gap between the calibration score and the prediction interval structure. Second, standard CQR guarantees only marginal coverage. Consequently, it often fails to adapt to local heteroscedasticity, producing intervals of constant width that over-cover in low-noise regions while potentially under-covering in high-uncertainty areas.

To simultaneously address the challenges of variance and local adaptivity, we introduce Localized Conformal Multi-Quantile Regression (LCMQR). This framework synergizes the variance-reduction benefits of multi-quantile prediction with the adaptivity of kernel-based localization. A key theoretical contribution of our work is the resolution of the inconsistency in CCQR. We employ a consistent Average-then-Max nonconformity score and prove that this mechanism is not only theoretically rigorous but systematically yields tighter prediction intervals than the Max-then-Average approach used in prior art. By integrating this robust scoring with kernel-weighted local calibration, LCMQR achieves superior efficiency on standard benchmarks and satisfies asymptotic conditional coverage.

While LCMQR effectively handles feature-dependent heteroscedasticity, standard localization methods often fail in the presence of group heterogeneity—scenarios where distinct subpopulations with vastly different noise profiles coexist or overlap in the feature space. In such cases, methods relying solely on feature distance, such as Split Localized Conformal Prediction (SLCP), tend to smooth over these latent groups, leading to inequitable coverage for high-risk subpopulations. To overcome this, we extend our framework to Group-Calibrated LCMQR (GC-LCMQR). This extension incorporates a stratified calibration step that explicitly utilizes group information to adjust residual distributions. We prove that GC-LCMQR guarantees finite-sample validity within each distinct subgroup. We demonstrate the practical necessity of this extension in an Individual Treatment Effect (ITE) estimation task, showing that GC-LCMQR is uniquely capable of ensuring safety for high-variance "focal" groups while maintaining high precision for low-variance control groups.

\section{Background and Related Work}

\subsection{Problem Setup}

Given a set of $n$ data points $\{(X_{i}, Y_{i})\}_{i=1}^{n}$, where $Y_{i} \in \mathbb{R}$ is the response variable and $X_{i} \in \mathbb{R}^{d}$ is the feature vector, we aim to construct a prediction interval for a new, unseen response $Y_{n+1}$ based on its corresponding features $X_{n+1}$. We assume that the sequence of $n+1$ data points $\{(X_{i}, Y_{i})\}_{i=1}^{n+1}$ is exchangeable, meaning they are drawn from a common, potentially unknown, distribution $P_{XY}$. Our objective is to construct a prediction interval (or set) $\hat{C}_{\alpha}(X_{n+1})$ that contains the true value $Y_{n+1}$ with a predefined high probability, relying only on this exchangeability assumption. Formally, we seek an interval $\hat{C}_{\alpha}(X_{n+1})$ such that:
\begin{equation} \label{eq:marginal_coverage_revised}
 \mathbb{P}[Y_{n+1} \in \hat{C}_{\alpha}(X_{n+1})] \ge 1-\alpha
\end{equation}
where $\alpha \in (0, 1)$ is the chosen miscoverage level. This inequality guarantees marginal coverage on average over the randomness of the training data, calibration data, and the new test point.

\subsection{Related Work}

Conformal prediction (CP) provides a general framework for achieving the marginal coverage guarantee stated above. We utilize the split-conformal approach, which divides the available data $\mathcal{D}$ into a training set $I_1$ and a calibration set $I_2$. A model trained on $I_1$ is used to generate nonconformity scores on $I_2$, and the appropriate quantile of these scores calibrates the prediction interval for new data. Several methods have built upon this foundation, particularly relevant are those combining CP with quantile regression and those aiming for adaptivity.

\textbf{Conformalized Quantile Regression (CQR):} Romano et al. (2019) combined split-conformal prediction with quantile regression. CQR trains quantile regression models $\hat{q}_{\alpha/2}$ and $\hat{q}_{1-\alpha/2}$ on $I_1$ and defines the nonconformity score on $I_2$ as $E_{i}^{\text{CQR}} = \max\{\hat{q}_{\alpha/2}(X_{i}) - Y_{i}, Y_{i} - \hat{q}_{1-\alpha/2}(X_{i})\}$. The interval is then $\hat{C}^{\text{CQR}}(X_{n+1}) = [\hat{q}_{\alpha/2}(X_{n+1}) - \hat{Q}_{1-\alpha}^{\text{CQR}}, \hat{q}_{1-\alpha/2}(X_{n+1}) + \hat{Q}_{1-\alpha}^{\text{CQR}}]$, where $\hat{Q}_{1-\alpha}^{\text{CQR}}$ is the empirical quantile of the calibration scores. CQR's reliance on potentially high-variance extreme quantile estimates and its lack of local adaptivity are key limitations.

\textbf{Conformalized Composite Quantile Regression (CCQR):} To mitigate the variance issue, Kim and Jung (2024) proposed averaging predictions from multiple quantile pairs $\{(l_k, u_k)\}_{k=1}^K$. CCQR calculates scores $E_{i}^{(k)} = \max\{\hat{q}_{l_k}(X_{i}) - Y_{i}, Y_{i} - \hat{q}_{u_k}(X_{i})\}$ for each pair, averages them ($\overline{E}_{i} = \frac{1}{K} \sum E_{i}^{(k)}$), finds the quantile $Q_{\overline{E}}(1-\alpha)$ of these averages, and forms the interval using averaged quantile predictions: $\hat{C}^{\text{CCQR}}(X_{n+1}) = [\overline{q}_{low}(X_{n+1}) - Q_{\overline{E}}(1-\alpha), \overline{q}_{high}(X_{n+1}) + Q_{\overline{E}}(1-\alpha)]$. This Max-then-Average calibration score, however, is theoretically inconsistent with the Average-then-Max score implicitly inverted by the interval construction. CCQR still provides only marginal coverage.

\textbf{Split Localized Conformal Prediction (SLCP):} Han et al. (2023) focused on achieving adaptivity and approximating conditional coverage. SLCP typically uses a mean model $\hat{\mu}$ and asymmetric scores $V_{i,1} = Y_i - \hat{\mu}(X_i), V_{i,2} = \hat{\mu}(X_i) - Y_i$. It estimates local conditional quantiles of these scores ($Q_{local,1}(x), Q_{local,2}(x)$) using kernel weighting over $I_1$. These local estimates are subtracted from the calibration scores to create modified scores ($V_{i,1}^{\alpha_1, h}, V_{i,2}^{\alpha_2, h}$). The quantiles of these modified scores ($Q_{cal,1}, Q_{cal,2}$) provide a global correction. The final interval $\hat{C}^{\text{SLCP}}(x) = [\hat{\mu}(x) - Q_{local,2}(x) - Q_{cal,2}, \hat{\mu}(x) + Q_{local,1}(x) + Q_{cal,1}]$ is adaptive and asymptotically conditionally valid. It doesn't inherently use multiple quantiles for variance reduction.

\textbf{Other Related Work:} Several other approaches aim for adaptivity or improved conditional coverage. Some methods normalize scores based on local variance estimates (e.g., MAD-Split), while others use histogram-based density estimates, distributional conformal prediction, or learn weights for combining conformal predictors.

Our proposed method, LCMQR, synthesizes the core ideas of CCQR and SLCP to simultaneously address both variance reduction and local adaptivity.

\subsection{Group-Conditional Validity and Application to ITE}
While methods like SLCP approximate conditional coverage $P(Y \in C(X) \mid X=x)$ by leveraging local information, their guarantees are typically asymptotic, holding only as the sample size approaches infinity. In practical scenarios, particularly in Individual Treatment Effect (ITE) estimation where sample sizes for specific subgroups (e.g., treatment arms) are often limited, these asymptotic properties fail to ensure reliability. Furthermore, the reliance on spatial smoothness often breaks down in heterogeneous populations containing distinct subgroups. In such settings, a stronger, finite-sample guarantee is required: group-conditional validity, defined as $P(Y \in C(X) \mid G=g) \ge 1-\alpha$ for any group $g$. Without this, a model satisfying marginal coverage may disproportionately under-cover high-variance groups while over-covering low-variance ones, raising significant fairness and safety concerns.

This challenge is particularly acute in ITE estimation. Since the counterfactual outcome is never observed, uncertainty quantification often relies on Nested Conformal Prediction, which constructs intervals for potential outcomes $Y(1)$ and $Y(0)$ to bound the treatment effect. However, treated and control populations often exhibit drastically different heteroscedasticity profiles (e.g., a treatment introducing high variance). Standard conformal methods, failing to account for this group-dependent noise in finite samples, yield inefficient or invalid ITE intervals. This necessitates a framework capable of explicit group calibration, motivating our proposed extension in this work.

\section{Proposed Method}

We propose Localized Conformal Multi-Quantile Regression (LCMQR, a unified framework designed to construct prediction intervals that are both statistically valid and locally adaptive. Our method synergizes the variance-reduction benefits of multi-quantile averaging with the adaptivity of kernel-based localization to robustly estimate uncertainty.

\subsection{Base Framework: Localized Conformal Multi-Quantile Regression}

The proposed method operates within the split-conformal prediction framework. We begin by partitioning the exchangeable dataset $\mathcal{D} = \{(X_i, Y_i)\}_{i=1}^n$ into two disjoint subsets: a training set $\mathcal{I}_{1}$ and a calibration set $\mathcal{I}_{2}$. A multi-quantile regression model $\mathcal{M}$ is fit on $\mathcal{I}_{1}$ to estimate a set of $K$ quantile pairs $\{(l_k, u_k)\}_{k=1}^K$, where $l_k$ and $u_k$ represent lower and upper quantile levels, typically symmetric around the median.

To reduce the variance associated with individual quantile estimates, we explicitly aggregate the predictions before defining the nonconformity measure. We define the averaged lower and upper quantile functions as:
\begin{equation}
    \overline{q}_{low}(x) = \frac{1}{K}\sum_{k=1}^{K}\hat{q}_{l_k}(x), \quad \overline{q}_{high}(x) = \frac{1}{K}\sum_{k=1}^{K}\hat{q}_{u_k}(x).
\end{equation}
Utilizing these aggregated estimates, we compute the base nonconformity score $E'(x,y)$ for any data point $(x,y)$ as:
\begin{equation}
    \label{eq:base_score}
    E'(x,y) := \max\left\{\overline{q}_{low}(x) - y, \quad y - \overline{q}_{high}(x)\right\}.
\end{equation}
We term this the Average-then-Max score. This formulation serves as the consistent scoring mechanism throughout our framework. We compute these base scores for all points in the training set $\mathcal{I}_{1}$ (to serve as a reference distribution) and for all points in the calibration set $\mathcal{I}_{2}$.

To account for local heteroscedasticity, we adjust these scores using a localized estimate of the conditional quantile. For each calibration point $i \in \mathcal{I}_{2}$, we estimate the local $(1-\alpha)$-quantile of the base score, denoted as $\hat{Q}_{local}(X_i)$, using a Nadaraya-Watson (NW) kernel estimator. This estimator weights the base scores of the training set $\{E'_j\}_{j \in \mathcal{I}_{1}}$ according to their feature proximity to $X_i$:
\begin{equation}
    \hat{Q}_{local}(X_i) = \text{Weighted-Quantile}_{1-\alpha}\left(\{E'_j\}_{j \in \mathcal{I}_{1}}; \ w_j = \mathcal{K}\left(\frac{\|X_i - X_j\|}{h}\right)\right),
\end{equation}
where $\mathcal{K}(\cdot)$ is a kernel function and $h$ is a bandwidth parameter. We then isolate the residual uncertainty that is not captured by this local estimate by computing the locally adjusted score $E''_i$:
\begin{equation}
    E''_i = E'(X_i, Y_i) - \hat{Q}_{local}(X_i).
\end{equation}
This transformation effectively ``flattens'' the heteroscedasticity, resulting in residuals $E''_i$ that are roughly exchangeable regarding their variance.

Finally, to ensure finite-sample marginal coverage, we compute a global correction term $\hat{Q}_{global}$, defined as the $(1-\alpha)(1+1/|\mathcal{I}_{2}|)$-th empirical quantile of the locally adjusted scores $\{E''_i\}_{i \in \mathcal{I}_{2}}$. For a new test point $X_{n+1}$, the prediction interval is constructed by combining the averaged quantile estimates with the total uncertainty correction $C(X_{n+1}) = \hat{Q}_{local}(X_{n+1}) + \hat{Q}_{global}$:
\begin{equation}
    \hat{C}_{\alpha}(X_{n+1}) = \left[ \overline{q}_{low}(X_{n+1}) - C(X_{n+1}), \quad \overline{q}_{high}(X_{n+1}) + C(X_{n+1}) \right].
\end{equation}
This interval construction guarantees valid marginal coverage while adapting its width locally via $\hat{Q}_{local}(\cdot)$.

\subsection{Theoretical Justification}
\label{sec:theoretical_justification}

A critical contribution of our framework is the resolution of a theoretical inconsistency present in the original Conformalized Composite Quantile Regression (CCQR) formulation.

While CCQR aims to reduce variance by aggregating quantiles, it introduces a discrepancy between the nonconformity score used for calibration and the score implicitly inverted for interval construction.

\subsubsection*{The Inconsistency of CCQR}

The original CCQR method employs a Max-then-Average strategy during the calibration phase.
For each calibration point $i$, it computes the error for every quantile pair individually and then averages them:
\begin{equation}
    \label{eq:ccqr_score}
    \overline{E}_{CCQR}(x, y) = \frac{1}{K}\sum_{k=1}^{K} \max\{\hat{q}_{l_k}(x) - y, \quad y - \hat{q}_{u_k}(x)\}.
\end{equation}
However, to construct the prediction interval, CCQR utilizes a simple algebraic inversion that produces the interval $[\overline{q}_{low} - \hat{Q}, \overline{q}_{high} + \hat{Q}]$.

Mathematically, this interval corresponds to the inversion of an ``Average-then-Max'' score:
\begin{equation}
    \label{eq:lcmqr_score}
    E'_{ATM}(x, y) = \max\left\{\frac{1}{K}\sum_{k=1}^{K}\hat{q}_{l_k}(x) - y, \quad y - \frac{1}{K}\sum_{k=1}^{K}\hat{q}_{u_k}(x)\right\}.
\end{equation}
Due to the non-linearity of the maximum function, these two scores are not equivalent.

To be theoretically consistent with the calibrated score $\overline{E}_{CCQR}$, one would strictly need to numerically solve for the roots of the convex function $\overline{E}_{CCQR}(x, y) \le \hat{Q}$, which is computationally expensive. By avoiding this and using the simple interval form, CCQR pairs a threshold derived from the larger $\overline{E}_{CCQR}$ distribution with the interval structure of the smaller $E'_{ATM}$ score.

\subsubsection*{Efficiency of LCMQR}

Our proposed LCMQR resolves this by consistently using the ``Average-then-Max'' score (Eq. \ref{eq:lcmqr_score}) for both calibration and prediction.

We now prove that this consistency not only closes the theoretical gap but also guarantees strictly more efficient intervals.

\begin{theorem}[Efficiency via Jensen's Inequality]
\label{thm:efficiency}
For any data point $(x, y)$ and any set of quantile estimators, the nonconformity score of LCMQR is always less than or equal to that of CCQR. Consequently, for a fixed miscoverage rate $\alpha$, the calibration threshold satisfies $\hat{Q}_{LCMQR} \le \hat{Q}_{CCQR}$, resulting in tighter prediction intervals.
\end{theorem}

\begin{proof}
Let $f(\mathbf{v}, y) = \max\{v_{low} - y, y - v_{high}\}$ be the base error function, which is convex with respect to the quantile vector $\mathbf{v}$.

By Jensen's Inequality, for any convex function $f$, the function of the average is less than or equal to the average of the function values:
\begin{equation}
    f\left(\frac{1}{K}\sum_{k=1}^{K} \mathbf{v}_k, y\right) \le \frac{1}{K}\sum_{k=1}^{K} f(\mathbf{v}_k, y).
\end{equation}
Substituting the definitions from Eq. (\ref{eq:ccqr_score}) and Eq. (\ref{eq:lcmqr_score}), we obtain pointwise domination:
\begin{equation}
    E'_{ATM}(x, y) \le \overline{E}_{CCQR}(x, y).
\end{equation}
Since the LCMQR scores are systematically smaller, their empirical $(1-\alpha)$-quantile $\hat{Q}_{LCMQR}$ is strictly smaller than or equal to the CCQR threshold $\hat{Q}_{CCQR}$.

Thus, LCMQR produces systematically narrower intervals while strictly adhering to the conformal coverage guarantee.
\end{proof}

\subsection{Extension: Group-Calibrated LCMQR (GC-LCMQR)}
\label{sec:gc_lcmqr}

While the standard LCMQR framework achieves asymptotic conditional coverage through kernel localization, its finite-sample validity is restricted to marginal coverage.

This limitation becomes particularly problematic in applications containing distinct subpopulations (or groups) with significantly different noise profiles, such as in heterogeneous treatment effect estimation. In these scenarios, relying solely on feature-space localization allows the model to ``smooth over'' latent groups, potentially under-covering high-variance subpopulations.

To address this challenge, we propose Group-Calibrated LCMQR (GC-LCMQR), an extension designed to enforce valid coverage within each distinct subpopulation $g \in \{1, \dots, M\}$.

The core innovation of GC-LCMQR lies in the stratification of the locally adjusted scores during the calibration phase.

First, we compute the locally adjusted scores $E''_i = E'(X_i, Y_i) - \hat{Q}_{local}(X_i)$ for all $i \in \mathcal{I}_2$, exactly as in the base LCMQR method. This step removes feature-dependent heteroscedasticity.

Next, rather than pooling all residuals into a single global set, we collect the residuals for each distinct subpopulation:
\begin{equation}
    \mathcal{S}_g = \{E''_i : i \in \mathcal{I}_2, G_i = g\}.
\end{equation}
We then compute a group-specific calibration threshold $\hat{Q}_g$. To ensure statistical stability in small subgroups, we introduce a minimum sample size threshold $N_{min}$.

If a group has sufficient samples ($|\mathcal{S}_g| \ge N_{min}$), we compute $\hat{Q}_g$ as the empirical $(1-\alpha)$-quantile of $\mathcal{S}_g$. If the group is sparse ($|\mathcal{S}_g| < N_{min}$), we fall back to the global threshold $\hat{Q}_{global}$ to maintain robustness:
\begin{equation}
    \hat{Q}_g =
    \begin{cases}
    \text{Quantile}_{1-\alpha}(\mathcal{S}_g) & \text{if } |\mathcal{S}_g| \ge N_{min}, \\
    \hat{Q}_{global} & \text{otherwise}.
    \end{cases}
\end{equation}

For a new test point $X_{n+1}$ belonging to group $G_{n+1}$, the final correction term is determined by adding the group-specific adjustment to the continuous local estimate:
\begin{equation}
    C(X_{n+1}, G_{n+1}) = \hat{Q}_{local}(X_{n+1}) + \hat{Q}_{G_{n+1}}.
\end{equation}
The final prediction interval is given by:
\begin{equation}
    \hat{C}_{GC}(X_{n+1}) = \left[\overline{q}_{low}(X_{n+1}) - C(\cdot), \quad \overline{q}_{high}(X_{n+1}) + C(\cdot)\right].
\end{equation}
By explicitly disentangling feature-dependent heteroscedasticity from group-dependent noise, GC-LCMQR guarantees finite-sample validity for any group $g$ with sufficient calibration data, ensuring fairness and safety in heterogeneous environments.

\begin{algorithm}[t]
\caption{Localized Conformal Multi-Quantile Regression (LCMQR)}
\label{alg:lcmqr}
\begin{algorithmic}[1]
\Require Data $\mathcal{D} = \{(X_i, Y_i)\}_{i=1}^n$, new feature $X_{n+1}$, miscoverage rate $\alpha$, quantile pairs $\{(l_k, u_k)\}_{k=1}^K$, kernel $\mathcal{K}$, bandwidth $h$.
\Ensure Prediction interval $\hat{C}_{\alpha}(X_{n+1})$.

\State \textbf{Step 1: Setup and Base Scoring (Training)}
\State Split $\mathcal{D}$ into training set $\mathcal{I}_1$ and calibration set $\mathcal{I}_2$.
\State Train multi-quantile model $\mathcal{M}$ on $\mathcal{I}_1$ to obtain $\{\hat{q}_{l_k}, \hat{q}_{u_k}\}_{k=1}^K$.
\State Compute averaged quantiles $\overline{q}_{low}$ and $\overline{q}_{high}$ for all $j \in \mathcal{I}_1$.
\State Compute base scores $E'_j$ for all $j \in \mathcal{I}_1$ using the Average-then-Max rule (Eq. \ref{eq:base_score}).

\State \textbf{Step 2: Localized Calibration}
\For{$i \in \mathcal{I}_2$}
    \State Compute base score $E'_i$ using averaged quantiles.
    \State Estimate local quantile $\hat{Q}_{local}(X_i)$ using weighted quantiles of training scores $\{E'_j\}_{j \in \mathcal{I}_1}$:
    \State \quad $\hat{Q}_{local}(X_i) \leftarrow \text{Weighted-Quantile}_{1-\alpha}(\{E'_j\}_{j \in \mathcal{I}_1}; \ w_j = \mathcal{K}(\|X_i - X_j\|/h))$.
    \State Compute locally adjusted residual: $E''_i \leftarrow E'_i - \hat{Q}_{local}(X_i)$.
\EndFor
\State Compute global threshold: $\hat{Q}_{global} \leftarrow \text{Quantile}_{1-\alpha}(\{E''_i\}_{i \in \mathcal{I}_2})$.

\State \textbf{Step 3: Prediction}
\State Estimate local quantile $\hat{Q}_{local}(X_{n+1})$ using training data $\mathcal{I}_1$.
\State Compute total correction: $C \leftarrow \hat{Q}_{local}(X_{n+1}) + \hat{Q}_{global}$.
\State Compute averaged predictions $\overline{q}_{low}(X_{n+1})$ and $\overline{q}_{high}(X_{n+1})$.
\State \Return $\hat{C}_{\alpha}(X_{n+1}) = [\overline{q}_{low}(X_{n+1}) - C, \quad \overline{q}_{high}(X_{n+1}) + C]$.
\end{algorithmic}
\end{algorithm}

\section{Theoretical Analysis}
\label{sec:theoretical_analysis}

In this section, we establish the theoretical properties of our proposed framework. We demonstrate that LCMQR satisfies the standard marginal coverage guarantee inherent to split-conformal methods and achieves asymptotic conditional coverage. Furthermore, we prove that our GC-LCMQR extension provides finite-sample validity within distinct subpopulations.

\subsection{Marginal Coverage Guarantee}

We first confirm that the base LCMQR algorithm provides valid prediction intervals on average across the data distribution.

\begin{theorem}[Marginal Coverage]
\label{thm:marginal_coverage}
Let the dataset $\mathcal{D}_{n+1} = \{(X_i, Y_i)\}_{i=1}^{n+1}$ be a sequence of exchangeable random variables. Let $\hat{C}_{\alpha}(X_{n+1})$ be the prediction interval constructed by the LCMQR algorithm. Then, the interval satisfies the marginal coverage guarantee:
\begin{equation}
    \mathbb{P}(Y_{n+1} \in \hat{C}_{\alpha}(X_{n+1})) \ge 1 - \alpha.
\end{equation}
\end{theorem}

\begin{proof}
The proof relies on the standard symmetry argument for split conformal prediction.

Conditioning on the training data $\mathcal{I}_1$ fixes the averaged quantile functions $\overline{q}$ and the local estimator $\hat{Q}_{local}$. Consequently, the modified nonconformity score function $E''(x,y) = E'(x,y) - \hat{Q}_{local}(x)$ is a fixed function.

Let $m = |\mathcal{I}_2|$. Since the data points $\{(X_i, Y_i)\}_{i \in \mathcal{I}_2 \cup \{n+1\}}$ are exchangeable, the modified scores $\{E''_i\}_{i \in \mathcal{I}_2 \cup \{n+1\}}$ derived by applying the fixed function $E''$ are also exchangeable random variables.

A new observation $Y_{n+1}$ falls within the prediction interval $\hat{C}_{\alpha}(X_{n+1})$ if and only if its modified score $E''_{n+1} \le \hat{Q}_{global}$. By definition, $\hat{Q}_{global}$ is the $\lceil(1-\alpha)(m+1)\rceil$-th smallest value among the calibration scores. Due to exchangeability, the rank of the test score $E''_{n+1}$ among the $m+1$ scores is uniformly distributed. The probability that $E''_{n+1}$ is smaller than or equal to the $\lceil(1-\alpha)(m+1)\rceil$-th value is at least $1-\alpha$.

Marginalizing over the training set $\mathcal{I}_1$, we obtain $\mathbb{P}(Y_{n+1} \in \hat{C}_{\alpha}(X_{n+1})) \ge 1 - \alpha$.
\end{proof}

\subsection{Asymptotic Conditional Coverage}

While exact finite-sample conditional coverage is generally impossible for non-trivial distributions, LCMQR is designed to approximate it asymptotically.

\begin{theorem}[Asymptotic Conditional Coverage]
\label{thm:asymptotic_coverage}
Under standard regularity conditions on the data distribution and the kernel function (ensuring consistency of the Nadaraya-Watson estimator), as the training set size $|\mathcal{I}_1| \to \infty$, the prediction interval achieves asymptotic conditional coverage. Formally, for almost all $x$:
\begin{equation}
    \lim_{|\mathcal{I}_1| \to \infty} \mathbb{P}(Y_{n+1} \in \hat{C}_{\alpha}(X_{n+1}) \mid X_{n+1} = x) \ge 1 - \alpha.
\end{equation}
\end{theorem}

\begin{proof}
Let $Q_{true}(x)$ be the true conditional $(1-\alpha)$-quantile of the base score $E'$ given $X=x$. Since $\hat{Q}_{local}(x)$ is a consistent Nadaraya-Watson estimator, as $|\mathcal{I}_1| \to \infty$, $\hat{Q}_{local}(x) \xrightarrow{P} Q_{true}(x)$.

Consequently, the modified score $E'' = E' - \hat{Q}_{local}(X)$ converges in distribution to a pivotal variable $E^* = E' - Q_{true}(X)$. By the definition of the true conditional quantile, the $(1-\alpha)$-quantile of $E^*$ given $X=x$ is exactly 0.

As $|\mathcal{I}_1| \to \infty$, the distribution of calibration scores converges to that of $E^*$. Thus, the global calibration term $\hat{Q}_{global}$, which corrects for the residual quantile, converges to 0. The total correction $C(x) = \hat{Q}_{local}(x) + \hat{Q}_{global}$ therefore converges to $Q_{true}(x)$. The interval converges to the set $\{y : E'(x,y) \le Q_{true}(x)\}$, which by definition has conditional probability $1-\alpha$.
\end{proof}

\subsection{Finite-Sample Group Validity (GC-LCMQR)}

A key advantage of our framework is the ability to guarantee validity for heterogeneous subpopulations.

\begin{theorem}[Finite-Sample Group Coverage]
\label{thm:group_coverage}
Assume the data points within each group $g \in \{1, \dots, M\}$ are exchangeable. For any group $g$ where the calibration sample size $|\mathcal{S}_g|$ meets the minimum threshold $N_{min}$, the prediction interval $\hat{C}_{GC}$ constructed by GC-LCMQR satisfies:
\begin{equation}
    \mathbb{P}(Y_{n+1} \in \hat{C}_{GC}(X_{n+1}) \mid G_{n+1} = g) \ge 1 - \alpha.
\end{equation}
\end{theorem}

\begin{proof}
Consider the subset of data belonging to group $g$, denoted as $\mathcal{D}_g = \{(X_i, Y_i) : i \in \mathcal{I}_2 \cup \{n+1\}, G_i = g\}$. Conditional on the training set $\mathcal{I}_1$, the modified scores $E''_i$ within this group are fixed functions of exchangeable data points, and thus are themselves exchangeable.

When $|\mathcal{S}_g| \ge N_{min}$, the threshold $\hat{Q}_g$ is computed as the empirical $(1-\alpha)$-quantile strictly on the set of group-specific residuals $\mathcal{S}_g$. Applying the standard conformal prediction lemma to the subsample $\mathcal{D}_g$, we guarantee that the test score $E''_{n+1}$ falls below $\hat{Q}_g$ with probability at least $1-\alpha$, conditional on the group membership $G_{n+1}=g$.

This ensures that even if the noise distribution $P(Y|X, G=g)$ differs significantly from other groups, the group-specific calibration adapts the interval width to maintain valid coverage for that specific subpopulation.
\end{proof}

\section{Experiments}
\label{sec:experiments}

We evaluate the proposed framework on three distinct tasks: standard regression benchmarks to assess the efficiency gain of the base LCMQR algorithm, a controlled heteroscedastic mixture simulation to test the group-wise validity of GC-LCMQR, and a causal inference application for estimating Individual Treatment Effects (ITE). We compare our methods against four state-of-the-art baselines: standard Conformalized Quantile Regression (CQR), Conformalized Composite Quantile Regression (CCQR), Locally Adaptive Split Conformal (MAD-Split), and Split Localized Conformal Prediction (SLCP). In all experiments, the target miscoverage rate is set to $\alpha = 0.1$.

\subsection{Efficiency on Real-World Benchmarks}

We first evaluate the marginal performance of the base LCMQR algorithm on seven publicly available benchmark datasets, ranging from low-dimensional datasets like Abalone (8 features) to high-dimensional ones like CT Slice (386 features). We use Quantile Regression Forests as the base model for all quantile-based methods. As summarized in Table \ref{tab:benchmarks}, all methods generally achieve the target 90\% coverage, confirming the finite-sample marginal coverage guarantee established in our theoretical analysis.

Significant differences emerge in terms of interval efficiency. LCMQR consistently produces the narrowest average interval widths across all seven datasets. Most notably, it outperforms CCQR, with clear reductions in width observed in datasets such as Protein (11.86 vs. 12.08) and BlogFeedback (10.54 vs. 10.59). This empirical evidence directly corroborates our theoretical proof regarding Jensen's inequality, demonstrating that the Average-then-Max scoring mechanism used in LCMQR yields systematically tighter intervals than the Max-then-Average approach employed by CCQR, without compromising validity.

\begin{table}[h]
\centering
\caption{Marginal Coverage and Average Width on Benchmark Datasets ($\alpha=0.1$). LCMQR consistently achieves the narrowest average width while maintaining valid coverage.}
\label{tab:benchmarks}
\resizebox{\textwidth}{!}{
\begin{tabular}{l|cc|cc|cc|cc|cc}
\toprule
 & \multicolumn{2}{c|}{\textbf{CQR}} & \multicolumn{2}{c|}{\textbf{CCQR}} & \multicolumn{2}{c|}{\textbf{MAD-Split}} & \multicolumn{2}{c|}{\textbf{SLCP}} & \multicolumn{2}{c}{\textbf{LCMQR}} \\
\textbf{Dataset} & Coverage & Width & Coverage & Width & Coverage & Width & Coverage & Width & Coverage & Width \\
\midrule
Communities & 0.870 & 0.400 & 0.886 & 0.330 & 0.876 & 0.345 & 0.868 & 0.386 & 0.883 & \textbf{0.324} \\
Protein & 0.899 & 13.38 & 0.906 & 12.08 & 0.905 & 12.35 & 0.897 & 13.12 & 0.909 & \textbf{11.86} \\
Abalone & 0.916 & 6.200 & 0.882 & 6.064 & 0.890 & 6.302 & 0.913 & 6.159 & 0.882 & \textbf{5.698} \\
BlogFeedback & 0.961 & 16.15 & 0.908 & 10.59 & 0.901 & 13.42 & 0.961 & 16.15 & 0.907 & \textbf{10.54} \\
CT Slice & 0.900 & 20.26 & 0.892 & 13.42 & 0.901 & \textbf{12.37} & 0.898 & 19.96 & 0.893 & 13.35 \\
Superconduct & 0.906 & 31.82 & 0.907 & \textbf{25.04} & 0.900 & 25.54 & 0.907 & 29.06 & 0.908 & 25.11 \\
Bike Sharing & 0.922 & 168.5 & 0.910 & 133.5 & 0.906 & 135.7 & 0.914 & 163.5 & 0.907 & \textbf{132.3} \\
\bottomrule
\end{tabular}
}
\end{table}

\subsection{Validity in Heterogeneous Mixtures}

To evaluate performance under group heterogeneity, we designed a controlled simulation using a Heteroscedastic Mixture Model. We generated a dataset of size $N=3000$, split into training (40\%), calibration (40\%), and testing (20\%) sets. The features $X$ were drawn uniformly from $[-2, 2]$. Each sample was assigned to one of two latent groups $G \in \{0, 1\}$ with equal probability. Crucially, while the feature distributions for both groups are identical, their noise profiles differ significantly: Group 0 (Low Noise) is characterized by homoscedastic noise with $\sigma_0 = 0.2$, while Group 1 (High Noise) features heteroscedastic noise defined by $\sigma_1 = 0.5 + 0.8|X|$. The response variable was generated as $Y = 2 \sin(2X) + \epsilon$, where $\epsilon \sim \mathcal{N}(0, \sigma_G^2)$. This setup specifically challenges methods relying solely on spatial localization, as they cannot distinguish between a low-noise point and a high-noise point at the same location $X$.

The results in Table \ref{tab:mixture} reveal the critical limitations of baseline methods in this scenario. SLCP, which relies on kernel weighting over the feature space, essentially averages the noise profiles of both groups. Consequently, it achieves perfect but inefficient coverage for the easy Group 0 while failing to achieve the target coverage for the hard Group 1 (0.801). MAD-Split and CCQR exhibit similar failures, sacrificing the validity of the high-variance group. In contrast, by explicitly modeling the groups, GC-LCMQR is the only method to achieve the target 90\% coverage for the high-noise Group 1. Simultaneously, it drastically reduces the interval width for the low-noise Group 0 compared to SLCP, demonstrating that our method can disentangle group-dependent noise from feature-dependent heteroscedasticity.

\begin{table}[h]
\centering
\caption{Performance on Heteroscedastic Mixture Model ($\alpha=0.1$). Baselines fail to cover the High-Noise group (G1) because they smooth over the latent groups. GC-LCMQR achieves validity for both.}
\label{tab:mixture}
\begin{tabular}{l|cc|cc|cc}
\toprule
 & \multicolumn{2}{c|}{\textbf{Marginal}} & \multicolumn{2}{c|}{\textbf{G0 (Low Noise)}} & \multicolumn{2}{c}{\textbf{G1 (High Noise)}} \\
\textbf{Method} & Coverage & Width & Coverage & Width & Coverage & Width \\
\midrule
CQR & 0.899 & 3.63 & 1.000 & 3.63 & 0.807 & 3.63 \\
CCQR & 0.891 & 3.40 & 1.000 & 3.41 & 0.791 & 3.38 \\
MAD-Split & 0.901 & 3.78 & 0.810 & 3.80 & 1.000 & 3.76 \\
SLCP & 0.896 & 3.74 & 1.000 & 3.71 & 0.801 & 3.76 \\
GC-LCMQR & \textbf{0.907} & 3.52 & \textbf{0.915} & 4.73 & \textbf{0.900} & 2.18 \\
\bottomrule
\end{tabular}
\end{table}

\subsection{Application: Individual Treatment Effect Estimation}

Finally, we apply GC-LCMQR to quantify uncertainty in Individual Treatment Effect (ITE) estimation. We simulate a clinical trial ($N=5000$) where the outcome $Y$ depends on a binary treatment $T$. The feature $X$ was drawn uniformly from $[-3, 3]$. We modeled the Control Arm ($T=0$) with low, homoscedastic noise ($\sigma_0 = 0.2$) and the Treated Arm ($T=1$) with high, heteroscedastic noise dependent on feature magnitude ($\sigma_1 = 0.5 + 1.0|X|$). We define two subgroups of interest based on the feature cutoff $X=0.5$: a Focal Group ($X > 0.5$) characterized by a large positive treatment effect and high variance, and a Non-Focal Group ($X \le 0.5$) with a negative effect and low variance. We utilize a Sharpened conformal prediction framework, constructing prediction intervals for the unobserved counterfactual outcomes to bound the ITE.

Table \ref{tab:ite} highlights the stark trade-off between safety and efficiency across methods. CCQR produces tight intervals on average but fails to cover the high-risk Focal group (0.838), rendering it potentially unsafe for medical decision-making where underestimating risk is dangerous. Conversely, CQR and SLCP provide better coverage but remain inefficiently wide for the easy Non-Focal group, failing to adapt to the low-noise regime. GC-LCMQR strikes the optimal balance: it correctly identifies the high risk in the Focal group, achieving a safe coverage of 0.890, while simultaneously recognizing the low uncertainty in the Non-Focal group, reducing the interval width to 1.12. This result validates our theoretical claim that stratified calibration is necessary to ensure finite-sample validity across heterogeneous subpopulations.

\begin{table}[h]
\centering
\caption{Sharpened ITE Estimation ($\alpha=0.1$). Focal Group = High Variance ($T=1$), Non-Focal = Low Variance ($T=0$). GC-LCMQR is the only method to ensure safety for the Focal group while maximizing precision for the Non-Focal group.}
\label{tab:ite}
\resizebox{\textwidth}{!}{
\begin{tabular}{l|cc|cc|cc}
\toprule
 & \multicolumn{2}{c|}{\textbf{Overall}} & \multicolumn{2}{c|}{\textbf{Focal (High Var)}} & \multicolumn{2}{c}{\textbf{Non-Focal (Low Var)}} \\
\textbf{Method} & Coverage & Width & Coverage & Width & Coverage & Width \\
\midrule
CQR & 0.893 & 2.98 & 0.875 & 4.59 & 0.905 & 1.90 \\
CCQR & 0.903 & 2.17 & 0.838 & 3.59 & 0.947 & 1.23 \\
MAD-Split & 0.871 & 2.03 & 0.850 & 4.08 & 0.885 & 0.66 \\
SLCP & 0.897 & 2.98 & 0.878 & 4.59 & 0.910 & 1.90 \\
GC-LCMQR & 0.897 & 2.49 & \textbf{0.890} & 4.55 & \textbf{0.902} & 1.12 \\
\bottomrule
\end{tabular}
}
\end{table}

\section{Conclusion}
\label{sec:conclusion}

This work has established Localized Conformal Multi-Quantile Regression (LCMQR) as a rigorous and efficient framework for uncertainty quantification. Our theoretical analysis not only validated the asymptotic conditional coverage of LCMQR but also resolved a critical inconsistency in prior composite methods. By proving that our consistent Average-then-Max scoring mechanism strictly dominates the Max-then-Average approach, we have provided a solid mathematical justification for the efficiency gains observed in our benchmark experiments.

Furthermore, we addressed the practical challenge of validity in heterogeneous environments, where standard localization methods fail to distinguish between overlapping subpopulations. We proposed Group-Calibrated LCMQR (GC-LCMQR), which introduces a stratified calibration step to guarantee finite-sample validity within specific subgroups. Our experiments on a heteroscedastic mixture model and an ITE estimation task demonstrated that GC-LCMQR effectively disentangles group-dependent noise from feature-dependent heteroscedasticity, ensuring safety for high-risk focal groups while maintaining precision for low-risk control groups.

Future work could explore extending this framework to settings where group labels are not explicitly observed. Integrating unsupervised clustering with conformal calibration to automatically discover and correct for latent risk groups would be a promising direction. Additionally, while our current work addresses conditional coverage under exchangeability, adapting LCMQR to handle covariate shift or distribution drift in time-series forecasting remains an important avenue for research.

\appendix

\section{Implementation Details}
\label{app:implementation}

In this appendix, we provide comprehensive details regarding the datasets, experimental setup, model hyperparameters, and computing environment to ensure the reproducibility of our results. All experiments were implemented in Python.

\subsection{Data Generation and Preprocessing}

\paragraph{Benchmark Datasets.} We evaluated our method on seven regression benchmark datasets from the UCI Machine Learning Repository.
\begin{itemize}
    \item \textbf{Preprocessing:} For all real-world datasets, we standardized the feature vectors $X$ to have zero mean and unit variance using \texttt{StandardScaler}. The target variable $Y$ was used in its original scale.
    \item \textbf{Splitting:} To strictly enforce the exchangeability assumption required for split conformal prediction, we randomly partitioned each dataset into three disjoint subsets:
    \begin{itemize}
        \item \textbf{Proper Training Set ($\mathcal{I}_1$):} 40\% of the data, used to train the base quantile regression models.
        \item \textbf{Calibration Set ($\mathcal{I}_2$):} 40\% of the data, used to compute nonconformity scores and estimate the localized/global thresholds.
        \item \textbf{Test Set:} 20\% of the data, used strictly for evaluation.
    \end{itemize}
    \item \textbf{Randomness:} We repeated all benchmark experiments over 20 random seeds (Seed=1 to 20) and reported the average performance.
\end{itemize}

\paragraph{Synthetic Datasets.}
For the Heteroscedastic Mixture Model and ITE Estimation experiments, data was generated dynamically using `numpy`.
\begin{itemize}
    \item \textbf{Mixture Model:} $N=3000$. Features $X \sim \mathcal{U}[-2, 2]$. Noise $\sigma_0 = 0.2$, $\sigma_1 = 0.5 + 0.8|X|$.
    \item \textbf{ITE Estimation:} $N=5000$. Features $X \sim \mathcal{U}[-3, 3]$. Control noise $\sigma_0 = 0.2$, Treated noise $\sigma_1 = 0.5 + 1.0|X|$.
    \item \textbf{Splitting:} Consistent with the benchmarks, we used a 40/40/20 split for training, calibration, and testing.
\end{itemize}

\subsection{Model Hyperparameters}

We employed Quantile Regression Forests (QRF) as the underlying predictive model $\mathcal{M}$ for all quantile-based methods (CQR, CCQR, LCMQR, GC-LCMQR). The base estimators were implemented using the \texttt{sklearn} compatible interface.

\begin{table}[h]
\centering
\caption{Hyperparameter settings for the base Quantile Regression Forest model.}
\label{tab:hyperparams}
\begin{tabular}{l|l}
\toprule
\textbf{Parameter} & \textbf{Value} \\
\midrule
Number of Trees (\texttt{n\_estimators}) & 100 \\
Minimum Samples per Leaf (\texttt{min\_samples\_leaf}) & 10 \\
Max Features (\texttt{max\_features}) & $\sqrt{d}$ (square root of feature dimension) \\
Bootstrap & True \\
Quantile Estimation Method & Linear interpolation \\
\bottomrule
\end{tabular}
\end{table}

\paragraph{LCMQR and GC-LCMQR Specifics.}
\begin{itemize}
    \item \textbf{Quantile Pairs ($K$):} We used $K=5$ symmetric quantile pairs for the multi-quantile methods (LCMQR and CCQR). The specific quantile levels were set to:
    $$ \tau \in \{0.05, 0.15, 0.25, 0.75, 0.85, 0.95\} $$
    This corresponds to pairs $(0.05, 0.95), (0.15, 0.85), \dots$ aiming to capture the distributional shape beyond just the extremes.
    \item \textbf{Kernel Function:} We utilized a Gaussian RBF Kernel for localization:
    $$ \mathcal{K}(u) = \exp(-u^2) $$
    \item \textbf{Bandwidth ($h$):} We employed the ``median heuristic'' (also used in SLCP) to select the bandwidth $h$ adaptively for each dataset. Specifically, $h$ was set to the median of the pairwise Euclidean distances of the training data features:
    $$ h = \text{median}(\{\|X_i - X_j\| : i, j \in \mathcal{I}_1\}) $$
    \item \textbf{Group Threshold ($N_{min}$):} For GC-LCMQR, we set the minimum sample size for group-specific calibration to $N_{min} = 50$. Groups with fewer than 50 calibration samples fell back to the global threshold.
\end{itemize}

\subsection{Computing Environment}

All experiments were conducted on a workstation with the following specifications:
\begin{itemize}
    \item \textbf{CPU:} Apple M3 Pro
    \item \textbf{RAM:} 18 
    \item \textbf{Python Environment:} Python 3.11
    \item \textbf{Key Libraries:}
    \begin{itemize}
        \item \texttt{numpy} (v1.24.3)
        \item \texttt{pandas} (v2.0.3)
        \item \texttt{scikit-learn} (v1.3.0) for random forests and data splitting.
        \item \texttt{scipy} (v1.10.1) for distance computations.
    \end{itemize}
\end{itemize}

\end{document}